\documentstyle[12pt]{article}
\newcommand{\n}{\hspace*{-2.5mm}}
\newcommand{\gsim}{\;\rlap{\lower 3.5 pt \hbox{$\mathchar \sim$}} \raise 1pt
 \hbox {$>$}\;}
\newcommand{\lsim}{\;\rlap{\lower 3.5 pt \hbox{$\mathchar \sim$}} \raise 1pt
 \hbox {$<$}\;}
\newcommand{\re}{\mathop{\mbox{Re}}\nolimits}
\newcommand{\im}{\mathop{\mbox{Im}}\nolimits}
\newcommand{\tr}{\mathop{\mbox{tr}}\nolimits}
\newcommand{\arcosh}{\mathop{\mbox{arcosh}}\nolimits}
\newcommand{\arsinh}{\mathop{\mbox{arsinh}}\nolimits}
\newcommand{\di}{\mathop{\mbox{Li}_2}\nolimits}
\newcommand{\tri}{\mathop{\mbox{Li}_3}\nolimits}
\newcommand{\cld}{\mathop{\mbox{Cl}_2}\nolimits}
\newcommand{\clt}{\mathop{\mbox{Cl}_3}\nolimits}
\renewcommand{\O}{{\cal O}}

\begin{document}
\title{\vskip-3cm{\baselineskip14pt
\centerline{\normalsize\hfill MPI/PhT/94-69}
\centerline{\normalsize\hfill October 1994}
}
\vskip1.5cm
Two-Loop ${\cal O}(\alpha_s^2)$ Correction to the $H\to b\bar b$ Decay Rate
Induced by the Top Quark}
\author{Bernd A. Kniehl\\
Max-Planck-Institut f\"ur Physik, Werner-Heisenberg-Institut\\
F\"ohringer Ring 6, 80805 Munich, Germany}
\date{}
\maketitle
\begin{abstract}
We present the two-loop QED and QCD corrections to the $f\bar fH$ Yukawa
coupling that are induced by the exchange of a virtual photon or gluon,
respectively, with a heavy-fermion loop inserted.
As an application, we study the corresponding $\O(\alpha_s^2M_H^2/m_t^2)$
correction to the $H\to b\bar b$ decay rate.
\end{abstract}

A Higgs boson with $M_H\lsim135$~GeV decays dominantly to $b\bar b$ pairs
\cite{bak}.
This decay mode will be of prime importance for Higgs-boson searches at
LEP~2 \cite{gkw}, the Tevatron \cite{mar}---or a possible 4-TeV upgrade
thereof \cite{gun}---, and the next $e^+e^-$ linear collider \cite{imh}.
Techniques for the measurement of the $H\to b\bar b$ branching fraction at a
$\sqrt s=500$~GeV $e^+e^-$ linear collider have been elaborated in
Ref.~\cite{hil}.

The present knowledge of quantum corrections to the $H\to b\bar b$ decay
rate has been reviewed very recently in Ref.~\cite{bak}.
The full one-loop electroweak corrections to this process are well established
\cite{fle,hff}.
At two loops, the universal \cite{uni} and non-universal \cite{non}
$\O(\alpha_sG_Fm_t^2)$ corrections have recently been calculated.
The pure QCD corrections are most significant numerically.
In ${\cal O}(\alpha_s)$, their full $m_b$ dependence is known \cite{bra}.
In ${\cal O}(\alpha_s^2)$, the first \cite{gor} and second \cite{sur} terms
of the expansion in $m_b^2/M_H^2$ have been found.

The results of Refs.~\cite{gor,sur} take into account five quark flavours.
In $\O(\alpha_s^2)$, there are additional contributions involving a virtual
top quark, which are not formally suppressed.
These corrections may be viewed as the absorptive parts of the three-loop
Higgs-boson self-energy diagrams that involve a top-quark loop, a bottom-quark
loop, and two gluon lines.
These diagrams may be divided in two classes: the so-called double-triangle
diagrams and the the diagrams where a gluon line with a top-quark loop
insertion is attached in all possible ways to the one-loop seed diagram
involving the bottom quark.
The first class has been considered just recently \cite{ckk}.
In this letter, we shall study the second class.

We shall proceed along the lines of Ref.~\cite{kni},\footnote{We take this
opportunity to correct two misprints in the published version of
Ref.~\cite{kni}, which are absent in the preprint.
In the second line of Eq.~(5), the terms $-\varphi^2$ and 53/2 should be
replaced by $+\varphi^2$ and 53/3.}
where the two-loop
$f\bar f\gamma$ vertex correction due to a virtual massive fermion, $F$, was
derived in QED, assuming $m_f^2\ll |s|$, where $s$ is the photon invariant
mass squared.
In this limit, only the Dirac Form factor, which multiplies $\gamma^\mu$,
survives.
It has the form \cite{kni}
\begin{equation}
\label{dirac}
\Gamma^\mu=\gamma^\mu\left[1+{\alpha\over\pi}Q_f^2F_1(s)
+\left({\alpha\over\pi}\right)^2Q_f^2N_FQ_F^2F_2^{(F)}(s)+\cdots\right],
\end{equation}
where $Q_f$ is the electric charge of $f$ (in units of the positron charge),
$N_f=1$ (3) for leptons (quarks), and the dots represent other contributions
in $\O(\alpha^2)$ and higher orders.
By adjusting coupling constants and colour factors, one immediately obtains
the corresponding QCD expansion for the case when $f$ and $F$ are quarks.
Specifically, one substitutes $\alpha\to\alpha_s$,
$Q_f^2\to C_F=(N_c^2-1)/(2N_c)=4/3$, and $Q_F^2\to T=1/2$, where $N_c=3$ is
number of colours and $\tr t^at^b=T\delta^{ab}$, with $t^a$
$(a=1,\ldots,8)$ being the generators of the quark (defining) representation
of SU($N_c$).
In particular, the $m_t$-dependent $\O(\alpha_s^2)$ correction to the
non-singlet contribution to
$R=\sigma(e^+e^-\to\mbox{hadrons})/\sigma(e^+e^-\to\mu^+\mu^-)$ is given by
\begin{equation}
\label{rns}
\delta R_{NS}=2\left({\alpha_s\over\pi}\right)^2C_FT\re F_2^{(F)}(s),
\end{equation}
which is in agreement with Ref.~\cite{che}.

Similarly to Eq.~(\ref{dirac}), the QED expansion of the $f\bar fH$ coupling
may be written as
\begin{equation}
\label{scalar}
\Gamma=1+{\alpha\over\pi}Q_f^2H_1(s)
+\left({\alpha\over\pi}\right)^2Q_f^2N_FQ_F^2H_2^{(F)}(s)+\cdots.
\end{equation}
We have $H_1(s)=(1/4)\hat\delta_{em}|_{h=s}$, where
$\hat\delta_{em}$ may be found in Eq.~(2.18) of Ref.~\cite{hff}.
$H_1(s)$ is plagued by an infrared singularity, which is cancelled by a
similar contribution from soft-photon bremsstrahlung when a physical
observable, such as the $H\to f\bar f$ decay rate, is computed.
In Ref.~\cite{hff}, the infrared singularity is regularized by an
infinitesimal photon mass, $\mu$.
Up to terms proportional to $m_f^2/s$, one has
\begin{equation}
H_1(s)={L\over2}(z-1)-{z^2\over4}+{7\over2}\zeta(2)-{1\over2},
\end{equation}
where $\zeta(2)=\pi^2/6$, $L=\ln(\mu^2/m_f^2)$, and
$z=\ln(-s/m_f^2-i\epsilon)$.
In the following, we shall need an expression for $H_1(s)$ appropriate for a
neutral gauge boson with arbitrary mass, $\sqrt\sigma$.
{}From Eq.~(2.6) of Ref.~\cite{hff} one extracts
\begin{equation}
H_1(s,\sigma)={1\over2}\left(\di\left(1+{s+i\epsilon\over\sigma}\right)
-\zeta(2)\right),
\end{equation}
where $\di{}$ denotes the dilogarithm.

The Feynman diagrams contributing to $H_2^{(F)}(s)$ are depicted in
Fig.~\ref{one}.
$H_2^{(F)}(s)$ is infrared-finite and devoid of mass singularities associated
with $f$.
It may be conveniently calculated by convoluting $H_1(s,\sigma)$
with the imaginary part of the one-loop contribution of $F$ to the photon
self-energy,
\begin{equation}
\im\Pi_{AA}^{(F)}(s)={\alpha\over3}N_FQ_F^2sP(s),
\end{equation}
where
\begin{equation}
P(s)=\left(1+{2m_F^2\over s}\right)\sqrt{1-{4m_F^2\over s}}.
\end{equation}
The precise relation reads
\begin{equation}
H_2^{(F)}(s)={1\over3}\int_{4m_F^2}^\infty
{d\sigma\over\sigma}\,P(\sigma)H_1(s,\sigma).
\end{equation}
After a straightforward calculation, one finds
\begin{eqnarray}
\label{htwo}
3H_2^{(F)}(s)&\n=\n&\tri(-\rho_-^2)+{1\over3}\left(5-{1\over r}\right)
\sqrt{1+{1\over r}}\left(\di(-\rho_-^2)+\varphi^2+{\zeta(2)\over2}\right)
-{2\over3}\varphi^3-\zeta(2)\varphi\nonumber\\
&\n\n&\qquad{}+{2\over3}\left(-{14\over3}+{1\over r}\right)\gamma-\zeta(3)
+{82\over27}-{2\over3r},\qquad r\le-1,\nonumber\\
&\n=\n&\clt(2\Phi)-{1\over3}\left(5-{1\over r}\right)\sqrt{-{1\over r}-1}
\cld(2\Phi)\nonumber\\
&\n\n&\qquad{}+{2\over3}\left(-{14\over3}+{1\over r}\right)\gamma-\zeta(3)
+{82\over27}-{2\over3r},\qquad-1\le r\le0,\nonumber\\
&\n=\n&\tri(r_-^2)+{1\over3}\left(5-{1\over r}\right)\sqrt{1+{1\over r}}
\left(\di(r_-^2)+f^2-\zeta(2)\right)-{2\over3}f^3+2\zeta(2)f\nonumber\\
&\n\n&\qquad{}+{2\over3}\left(-{14\over3}+{1\over r}\right)g-\zeta(3)
+{82\over27}-{2\over3r}\nonumber\\
&\n\n&\qquad{}+i\pi\left[f^2-{1\over3}\left(5-{1\over r}\right)
\sqrt{1+{1\over r}}f+{14\over9}-{1\over3r}\right],\qquad r\ge0,
\end{eqnarray}
where $\zeta(3)=1.20205\,69031\,59594\,28540\ldots$, $\tri{}$ is the
trilogarithm, $\cld{}$ ($\clt{}$) is the (generalized) Clausen function of
second (third) order,
\begin{eqnarray}
r&\n=\n&{s\over4m_F^2},\qquad \rho_\pm=\sqrt{-r}\pm\sqrt{-r-1},\qquad
r_\pm=\sqrt{1+r}\pm\sqrt r,\nonumber\\
\varphi&\n=\n&\ln\rho_+=\arcosh\sqrt{-r},\qquad\Phi=\arcsin\sqrt{-r},\qquad
f=\ln r_+=\arsinh\sqrt r,\nonumber\\
\gamma&\n=\n&\ln(\rho_++\rho_-)=\ln\left(2\sqrt{-r}\right),\qquad
g=\ln(r_+-r_-)=\ln\left(2\sqrt r\right).
\end{eqnarray}
Note that $H_1(s)$ and $H_2^{(F)}(s)$ develop imaginary parts above the
$f\bar f$-pair production threshold, i.e., for $s>4m_f^2=0$.

It is instructive to study the limiting behaviour of $H_2^{(F)}(s)$.
For $s\to-\infty$ and $s\to-0$, one has
\begin{eqnarray}
\label{app}
3H_2^{(F)}(s)&\n=\n&-{2\over3}\gamma^3+{5\over3}\gamma^2
-\left(\zeta(2)+{28\over9}\right)\gamma-\zeta(3)+{5\over6}\zeta(2)+{82\over27}
+{3\gamma\over2r}\nonumber\\
&\n\n&\qquad{}+\O\left({\gamma^2\over r^2}\right),
\qquad r\ll-1,\nonumber\\
&\n=\n&{r\over5}\left(-4\gamma+{107\over15}\right)
+{r^2\over35}\left(6\gamma-{529\over70}\right)+\O(r^3\gamma),
\qquad-1\ll r\le0,
\end{eqnarray}
respectively.
The corresponding expansions for positive $s$ may be found by analytic
continuation, i.e., by substituting $\gamma=g-i\pi/2$.
In compliance with the Appelquist-Carazzone theorem \cite{app},
the loop fermion, $F$, decouples for $m_F^2\gg|s|$.

In Fig.~\ref{two}, $\re H_2^{(F)}(s)$ is plotted as a function of
$r=s/(4m_F^2)$.
At $r\approx5.62$, $\re H_2^{(F)}(s)$ assumes its maximum value, 0.432.
Its expansions, which emerge from Eq.~(\ref{app}) through analytic
continuation, are also shown.
Obviously, they provide an excellent approximation for $r\gsim1$ and $r\lsim1$,
respectively.

The QCD expansion of the $f\bar fH$ coupling for the case when $f$ and $F$
are quarks may be obtained from Eq.~(\ref{scalar}) through the substitutions
specified below Eq.~(\ref{dirac}).
As an application, we consider the $m_t$-dependent $\O(\alpha_s^2)$
correction to the $H\to b\bar b$ decay rate arising from the Feynman diagrams
in Fig.~\ref{one} with $f=b$, $F=t$, and the photon replaced by a gluon.
Similarly to Eq.~(\ref{rns}), the relative shift is
\begin{eqnarray}
\label{hqq}
{\delta\Gamma\left(H\to b\bar b\right)\over\Gamma\left(H\to b\bar b\right)}
&\n=\n&2\left({\alpha_s\over\pi}\right)^2C_FT\re H_2^{(t)}(M_H^2)\nonumber\\
&\n=\n&\left({\alpha_s\over\pi}\right)^2\left[{M_H^2\over45m_t^2}
\left(2\ln{m_t^2\over M_H^2}+{107\over15}\right)
+\O\left({M_H^4\over m_t^4}\ln{m_t^2\over M_H^2}\right)\right],
\end{eqnarray}
where the second line is appropriate for $M_H\lsim2M_W$, where the
$H\to b\bar b$ decay rate is most relevant phenomenologically.
The coefficient of $(\alpha_s/\pi)^2$ in Eq.~(\ref{hqq}) ranges between
$2.38\times10^{-2}$ for $(M_H,m_t)=(60,200)$~GeV and 0.502 for
$(M_H,m_t)=(1000,150)$~GeV; its value at $M_H=2m_t$ is 0.353.
This has to be compared with the value 29.14671 due to five massless quark
flavours \cite{gor}.
On the other hand, the finite-$m_b$ term in  $\O(\alpha_s^2)$ has the
coefficient $-87.72459\left(\overline m_b(M_H)/M_H\right)^2$ \cite{sur},
where $\overline m_b(\mu)$ is the bottom-quark $\overline{\mbox{MS}}$ mass at
renormalization scale $\mu$.
Assuming $\alpha_s(M_Z)=0.118$ \cite{bet} and $m_b=4.72$~GeV \cite{dom},
this amounts to $-0.200$ ($-2.37\times10^{-2}$) at $M_H=60$~GeV ($2M_W$).
In conclusion, the $\O(\alpha_s^2M_H^2/m_t^2)$ correction to
$\Gamma\left(H\to b\bar b\right)$ arising from the Feynman diagrams
shown in Fig.~\ref{one} is comparable in size with the
$\O(\alpha_s^2m_b^2/M_H^2)$ correction, but has the opposite sign.

\begin{figure}[p]

\centerline{\bf FIGURE CAPTIONS}

\caption{\label{one}Feynman diagrams pertinent to the two-loop $f\bar fH$
vertex correction induced by a virtual heavy fermion, $F$.}

\caption{\label{two}$\re H_2^{(F)}(s)$ as a function of $r=s/(4m_F^2)$
[see Eq.~(\protect\ref{htwo})] and its expansions
[see Eq.~(\protect\ref{app})].}

\end{figure}

\end{document}